\documentclass[11pt,letterpaper]{article}
\usepackage[margin=1.08in,top=.9in,bottom=.95in]{geometry}
\usepackage[T1]{fontenc}
\usepackage{lmodern,amsmath,amssymb,amsthm,microtype}
\usepackage[hidelinks]{hyperref}
\hypersetup{pdftitle={LogCFL = UAuxPDA(O(log n), poly(n))},
  pdfauthor={A. Pavan and N. V. Vinodchandran}}
\newcommand{\LogCFL}{\mathsf{LogCFL}}
\newcommand{\UAux}{\mathsf{UAuxPDA}}
\newcommand{\poly}{\operatorname{poly}}
\newtheorem*{theorem}{Theorem}
\newtheorem*{claim}{Claim}
\title{\(\LogCFL = \UAux\bigl(O(\log n),\poly(n)\bigr)\)}
\author{A. Pavan \and N. V. Vinodchandran}
\date{}

\begin{document}
\maketitle
\begin{abstract}
We observe that the isolation technique introduced by Hoza in his proof of
\(\mathsf{NL}=\mathsf{UL}\) extends to give
\(\LogCFL=\UAux(O(\log n),\poly(n))\).
\end{abstract}

\section{Introduction}
We observe that Hoza's technique~\cite{Hoza} extends to the complexity class \(\LogCFL\) and show that \(\LogCFL\) can be made unambiguous. We use notation from Hoza's paper~\cite{Hoza}. 

\section{Proof}
\begin{theorem}
Every language in \(\LogCFL\) is accepted by an unambiguous auxiliary pushdown
automaton in polynomial time with \(O(\log n)\) work space. That is, \(\LogCFL = \UAux\bigl(O(\log n),\poly(n)\bigr)\)
\end{theorem}

\begin{proof}
By Venkateswaran's characterization of \(\LogCFL\) as logspace-uniform
\(\mathsf{SAC}^1\)~\cite{Venkateswaran}, it suffices to evaluate a layered
semi-unbounded Boolean circuit of polynomial size \(N\) and depth
\(D=O(\log N)\). A semi-unbounded circuit has AND gates with fan-in 2 and OR gates with unbounded fan-in. Leaves are in level zero and wires join consecutive levels.
Fix the circuit and its input. A certificate tree for an accepting input chooses one true child at
each OR and chooses all children (at most two) at each AND gate.
Given a weight function $w$ that assigns weights to gates, the weight of a certificate tree $T$ is the sum
$W(T;w)=\sum_{x\in V(T)} w(g_x)$, where \(g_x\) is the circuit gate at
node \(x\), and $V(T)$ is the vertex set of the tree $T$.  We call $w$ \emph{min-unique} on this input if every true
gate has exactly one minimum-weight certificate tree rooted at it.

By the work of Reinhardt and Allender~\cite[Section~3]{RA}, it suffices
for unambiguous circuit evaluation to construct a logspace-computable,
polynomially bounded weight function that is min-unique for the given
circuit and input. However, we design a family of weight functions as opposed to a single weight function. Reinhardt and Allender also give an unambiguous auxiliary pushdown automaton that checks whether a given weight function is min-unique or not. Their work uses weight on circuit wires; however, van Melkebeek and Prakriya~\cite{vMP} design a similar procedure to check min-uniqueness when the weights are assigned to nodes.
We prove that for a given circuit and input, at least three-fourths of the weight functions are min-unique. By using the unambiguous procedure of Reinhardt and Allender and van Melkebeek and Prakriya, we can cycle through each weight function in the family till we find a min-unique weight function.

%We construct a polynomial-size
%family of candidate weight functions, enumerable in logspace. For every circuit
%of that size and every input, at least three quarters of the weight functions are
min-unique. Reinhardt and Allender also give an unambiguous auxiliary pushdown automaton that checks whether a given weight function is min-unique or not. Their work uses weight on circuit wires; however, van Melkebeek and Prakriya~\cite{vMP} design a similar procedure to check min-uniqueness when the weights are assigned to nodes.

\subsection*{Weight function}

 Let $N$ denote the number of gates in the circuit and let $D$ be its depth.   As in Hoza~\cite{Hoza}, let $E:[N] \rightarrow \{0, 1\}^\ell$ be a logspace-computable code with distance $\delta \ge\lceil\log_2(2N^2)\rceil$, where $\ell = O(\log N)$.

For a seed \(z\in\{0,1\}^{\ell}\) and a gate \(v\) at level \(h\), define
\[
 w_z(v)=1+\sum_{j=1}^{\ell}6^{\,h+\ell-j}E(v)_j z_j ,
\]

Since the height is $O(\log N)$, the weights are polynomial-bounded and each gate weight is computable in logspace.

Hoza uses the 2-adic valuation of the layer index to select a highest-priority
layer in each path interval. In our scenario, since the circuits are $\log$ depth, the weight function is slightly simpler. In a certificate tree the root is the unique
node at its highest level, and at most \(2^r\) nodes occur \(r\) levels
below it. We can therefore scale weights at level \(h\) by \(6^h\).
Since \(D=O(\log N)\), even \(6^D\) is polynomial in \(N\). Note that each $z \in \{0, 1\}^\ell$ define a weight function. Since $\ell = O(\log N)$, we have a polynomial-size family of weight functions.

\subsection*{Isolation}
We use Hoza's bit-flipping argument~\cite[Theorem~2.4]{Hoza}. For a true gate \(v\), let \(F_v(z)=\min_T W(T;w_z)\), where the minimum ranges
over all certificate trees rooted at \(v\).
Let \((\partial_j F_v)(z)=F_v(z_{j\gets 1})-F_v(z_{j\gets 0})\), keeping
the other bits fixed.

\begin{claim}
For every true gate \(v\) at level \(h\), every seed \(z\), and every
\(j\in\{1,\ldots,\ell\}\), let \(B_j=6^{h+\ell-j}\). Then
\begin{equation}
 B_j E(v)_j \ \le\ (\partial_j F_v)(z)\ \le\
 B_j\bigl(E(v)_j+\tfrac12\bigr).
 \label{eq:slope}
\end{equation}
\end{claim}
\begin{proof}[Proof of the claim]
Fix a certificate tree \(T\) rooted at \(v\), and write
\(W_b(T)=W(T;w_{z_{j\gets b}})\). Turning bit \(j\) on adds \(B_jE(v)_j\)
at the root. At distance \(r\) there are at most \(2^r\) nodes, each gaining
at most \(B_j/6^r\). The descendants therefore gain between zero and
\(B_j\sum_{r\ge1}(2/6)^r=B_j/2\). Hence for every \(T\),
\[
 W_0(T)+B_jE(v)_j \le W_1(T) \le W_0(T)+B_jE(v)_j+B_j/2 .
\]
Taking the minimum over the same set of trees in all three terms and
subtracting \(\min_T W_0(T)\) gives \eqref{eq:slope}, since
\(\min_T W_b(T)=F_v(z_{j\gets b})\).
\end{proof}

\begin{claim}
For distinct true gates \(u,v\) on the same level and a uniformly random
seed \(z\in\{0,1\}^{\ell}\),
\begin{equation}
 \Pr_z\bigl[F_u(z)=F_v(z)\bigr]\le 2^{-\delta}.
 \label{eq:collision}
\end{equation}
\end{claim}

\begin{proof}
Fix all coordinates where $E(u)_j = E(v)_j$.  At least $\delta$ free coordinates
remain where the codewords disagree.  We show at most \emph{one} assignment of these
free coordinates can produce a tie.

Suppose $z$ is a tie, i.e., $F_u(z) = F_v(z)$.  Let $z'$ be a string that differ from $z$ in at aleast one of the free coordinates.  Consider a process that converts  $z$ to $z'$  by flipping the free coordinates them in increasing order.  Let $j$ be the first coordinate
flipped.  WLOG $E(u)_j = 1$ and $E(v)_j = 0$.  Then by the interval bound:
\begin{align*}
  \partial_j F_u &\;\geq\; B_j \cdot 1 = B_j,\\
  \partial_j F_v &\;\leq\; B_j \cdot \tfrac{1}{2} = \tfrac{B_j}{2}.
\end{align*}
So flipping $z_j$ moves $F_u - F_v$ by at least $B_j - B_j/2 = B_j/2$ away from
zero.  For all later flips $k > j$, the total correction is at most
\[
  \tfrac{3}{2}\sum_{k > j} B_k
  \leq \tfrac{3}{2} \cdot B_j \sum_{s=1}^{\ell} 6^{-s}
  \leq  \tfrac{3}{2} \cdot \frac{B_j}{5}
  = \frac{3B_j}{10} < \frac{B_j}{2}.
\]
Hence the gap cannot return to zero, so $z'$ is not a tie.  At most one assignment
of the free coordinates yields a tie; since there are $2^m \geq 2^\delta$ assignments,
$\Pr[\text{tie}] \leq 2^{-\delta}$.
\qed
\end{proof}

\begin{claim}
For every fixed circuit and input, at least \(3/4\) of the seeds
\(z\in\{0,1\}^{\ell}\) give a min-unique weight function \(w_z\).
\end{claim}
\begin{proof}[Proof of the claim]
By \eqref{eq:collision}, each pair of distinct true gates on the same level
has equal minimum weights with probability at most \(2^{-\delta}\).
There are at most \(\binom{N}{2}\) such unordered pairs across all levels.
Since \(2^{-\delta}\le 1/(2N^2)\), the union bound gives
\[
 \Pr_z[\text{some such pair ties}]
 \le \binom{N}{2}2^{-\delta}\le \tfrac14 .
\]
Fix a seed with no such tie. We prove by induction on the level that every
true gate has a unique minimum-weight certificate tree. A true leaf has
only the one-node certificate. At a true OR gate, all true children lie on
the preceding level and have distinct minimum weights. A minimum tree must
choose the child \(u\) with the smallest \(F_u(z)\) and its unique minimum tree. At a true AND
gate, a minimum tree must use a minimum tree for every child. Each is
unique by induction, so their combination in the prescribed order is unique.
Thus every seed with no tie is min-unique, proving the claim.
\end{proof}

In particular, the family contains the required min-unique weighting.
\end{proof}

\section*{AI disclosure}
GPT Astra 6.2 found the extension of Hoza's isolation technique to \(\LogCFL\).  The authors take full responsibility for the contents of this note.

\end{document}